\documentclass[arxiv, preprint]{imsart}

\usepackage{amsmath, amsthm, amssymb}

\usepackage{times}
\usepackage{bm}
\usepackage{natbib}
\usepackage{tablefootnote}
\usepackage[plain,noend]{algorithm2e}

\usepackage{changepage}

\usepackage[dvipsnames]{xcolor}

\def\by{y} 

\def\pr{\text{pr}}
\def\ep{E}
\def\var{\text{var}}       
\def\cov{\text{cov}}

\usepackage{graphicx, caption} 
\usepackage{multirow}

\theoremstyle{plain}

\newtheorem{theorem}{Theorem}[section]
\newtheorem{condition}[theorem]{Condition}
\theoremstyle{remark}

\newtheorem{remark}{Remark}

\usepackage{fullpage}

\begin{document}

\begin{frontmatter}
\title{Asymptotic distribution-free change-point detection for data with repeated observations}
\runtitle{Change-point detection for data with repeated observations}

\begin{aug}
\author{\fnms{Hoseung} \snm{Song}\ead[label=e1]{hosong@ucdavis.edu}}
\and
\author{\fnms{Hao} \snm{Chen}\ead[label=e2]{hxchen@ucdavis.edu}}
\runauthor{H. Song and H. Chen}
\address{University of California, Davis}
\end{aug}

\begin{abstract}
In the regime of change-point detection, a nonparametric framework based on scan statistics utilizing graphs representing similarities among observations is gaining attention due to its flexibility and good performances for high-dimensional and non-Euclidean data sequences, which are ubiquitous in this big data era. However, this graph-based framework encounters problems when there are repeated observations in the sequence, which often happens for discrete data, such as network data. In this work, we extend the graph-based framework to solve this problem by averaging or taking union of all possible optimal graphs resulted from repeated observations. We consider both the single change-point alternative and the changed-interval alternative, and derive analytic formulas to control the type I error for the new methods, making them fast applicable to large datasets. The extended methods are illustrated on an application in detecting changes in a sequence of dynamic networks over time. All proposed methods are implemented in an  $\texttt{R}$ package $\texttt{gSeg}$ available on CRAN.
\end{abstract}



\end{frontmatter}

\allowdisplaybreaks

\section{Introduction}  \label{sec:int}


\subsection{Background}
Change-point analysis plays a significant role in various fields when a sequence of observations is collected. In general, the problem concerns testing whether or not a change has occurred, or several changes might have occurred, and identifying the locations of any such changes. In this paper, we consider the offline change-point detection problem where a sequence of independent observations $\{\by_{i}\}_{i=1,\ldotp\ldotp\ldotp,n}$ is completely observed at the time when data analysis is conducted. Here, $n$ is the length of the sequence, and $i$ is the time index or other meaningful index depending on the specific application. We consider testing the null hypothesis
\begin{equation} \label{eq1.1}
H_{0} : \by_{i} \sim F_{0}, \ i = 1,\ldotp\ldotp\ldotp,n,
\end{equation}
against the single change-point alternative
\begin{equation} \label{eq1.2}
H_{1} : \exists \ 1\le \tau\le n, \ \by_{i} \sim \left\{
\begin{tabular}{c} 
$F_{1}$,  \ $i > \tau$ \ \ \ \  \ \ \ \\
$F_{0}$, \ otherwise, \\
\end{tabular}
\right.
\end{equation}
or the changed-interval alternative
\begin{equation} \label{eq1.3}
H_{2} : \exists \ 1\le \tau_{1}\le \tau_{2}\le n, \ \by_{i} \sim \left\{
\begin{tabular}{c} 
$F_{1}$, \  $i = \tau_{1}+1,\ldotp\ldotp\ldotp,\tau_{2}$  \\
$F_{0}$, \ otherwise, \ \ \ \ \ \ \ \  \ \ \ \ \\
\end{tabular}
\right.
\end{equation}
where $F_{0}$ and $F_{1}$ are two different distributions.

This problem has been extensively studied for univariate data; see \cite{chen2011parametric} for a survey. However, in many modern applications, $\by_i$ is high-dimensional or even non-Euclidean. For high-dimensional data, most methods are based on parametric models; see for example \citet{Zhang2010detecting, wang2017change, wang2018high}. To apply these methods, the data sequence needs to follow specific parametric models. In the nonparametric domain, \citet{harchaoui2009kernel} made use of kernel methods, \citet{lung2011homogeneity} utilized marginal rankings, and \citet{matteson2014nonparametric} made use of distances among observations.  These methods can be applied to a wider range of problems than parametric methods.  However, it is in general difficult to conduct theoretical analysis on nonparametric methods and none of these nonparametric methods provides analytic formulas for false discovery control, making them difficult to be applied to large datasets.

\subsection{Graph-based change-point methods} \label{restriction}

Recently, \citet{chen2015graph}  and \citet{chu2019asymptotic} developed a graph-based framework to detect the change-point for high-dimensional and non-Euclidean data sequences. This framework is based on a similarity graph $G$, such as a minimum spanning tree (MST),  which is a spanning tree that connects all observations with the sum of distances of the edges in the tree minimized, constructed on the sample space. Based on $G$ over $t$, test statistics rely on three basic quantities, $R_{0,G}(t)$, $R_{1,G}(t)$, and $R_{2,G}(t)$, where for each $t$ $R_{0,G}(t)$ is the number of edges connecting observations before and after $t$, $R_{1,G}(t)$ is the number of edges connecting observations prior to $t$, and $R_{2,G}(t)$ is the number of edges connecting observations after $t$. Then, four scan statistics were studied: the original edge-count scan statistic $Z(t)$, the weighted edge-count scan statistic $Z_{w}(t)$, the generalized edge-count scan statistic $S(t)$, and the max-type edge-count scan statistic $M(t)$ that can be applied to various alternatives. For detailed comparisons, see \cite{chu2019asymptotic}.

While the methods proposed by \citet{chen2015graph} and \citet{chu2019asymptotic} work well for continuous data, they are problematic for data with repeated observations, which is common for discrete data, such as network data.  The reason is that these methods depend on the similarity graph constructed on observations. When there are repeated observations, the similarity graph is in general not uniquely defined and these methods are troublesome. For example, \citet{chen2015graph} analyzed a phone-call network dataset and the task was to test whether there is any change in the dynamics of the networks over time.  In this dataset, a few networks in the sequence are exactly the same. \citet{chen2015graph} used the MST as the similarity graph. 
More specifically, in the phone-call network dataset, there are in total 330 networks $\{\by_{1}, \ldots, \by_{330}\}$ in the sequence and among them are 290 distinct networks. For example, $\by_{1}$, $\by_{6}$, $\by_{16}$ are exactly the same. All repeated observations are listed in Supplement A. Due to this, there are numerous ways in assigning edges in the MST with repeated observations. Hence, the MST is not uniquely defined and existing graph-based methods are not reliable since they are formulated by the unique similarity graph on pooled observations.
\begin{table*} [h]
	\caption{The $p$-values and corresponding test statistics (in parentheses) for four testing procedures proposed in \citet{chen2015graph} and \citet{chu2019asymptotic}: an original edge-count scan statistic $(\max_{n_0\leq t\leq n_1} Z_0(t))$, a generalized edge-count scan statistic $(\max_{n_0\leq t\leq n_1} S(t))$, a weighted edge-count scan statistic $(\max_{n_0\leq t\leq n_1} Z_w(t))$, and a max-type edge-count scan statistic $(\max_{n_0\leq t\leq n_1} M(t))$.  Here, $n_0$ is set to be $\lceil0.05n\rceil$  = 17  and $n_{1} = 330-n_{0}$, where $\lceil x\rceil$ denotes the smallest integer greater than or equal to $x$.}
	\label{tab:phone}
	\centering
	\begin{tabular}{clll}
		& MST \#1 & MST \#2 & MST \#3  \\
		$\max_{n_0\leq t\leq n_1} Z_0(t)$ & 0.09 (2.32)& 0.91 (0.92) & 0.51 (1.57) \\
		$\max_{n_0\leq t\leq n_1} S(t)$ & 0.04 (13.61)& 0.08 (12.31) & 0.01 (16.36) \\
		$\max_{n_0\leq t\leq n_1} Z_w(t)$ & 0.44 (2.11) & 0.02 (3.49) & 0.88 (1.54)  \\
		$\max_{n_0\leq t\leq n_1} M(t)$ & 0.09 (3.05) & 0.02 (3.49) & 0.05 (3.27)  \\
	\end{tabular}
\end{table*}

Table \ref{tab:phone} lists test statistics and their corresponding $p$-values of four testing procedures proposed in \citet{chen2015graph} and \citet{chu2019asymptotic} on three randomly chosen MSTs. We see that the $p$-value depends heavily on the choice of the MST: It could be very small under one MST, but very large on another MST, leading to completely different conclusions on whether the sequence is homogeneous or not. Moreover, since the number of possible MSTs is huge, it is impractical to compute the test statistics on all possible MSTs directly to get a summary.


\subsection{Our contribution}

We extend the methods in \cite{chen2013graph} and \cite{zhang2017graph} to the change-point settings and propose new graph-based testing procedures that can deal with repeated observations properly. This work fills the gap for the graph-based framework in dealing with discrete data. We show that the new tests are asymptotically distribution-free under the null hypothesis of no change and reveal that the limiting distributions for two approaches in \cite{chen2013graph} are the same, even for continuous data. We also derive analytic formulas to approximate permutation $p$-values for those modified test statistics, making them fast applicable to real datasets. To improve the analytical $p$-value approximations for finite sample sizes, skewness correction is also performed. We show that the proposed tests work well to detect the change when the data has repeated observations. We illustrate the new testing procedures through an analysis on phone-call network dataset. The new methods are implemented in an  $\texttt{R}$ package $\texttt{gSeg}$ {available on CRAN}.



\section{Notations and related existing works} \label{not}

For data with repeated observations, we represent the data using a contingency table for each $t$. Suppose that there are in total $n$ observations and $K$ distinct values, which we also refer to as categories in the following. Each $t$ divides the sequence into two groups, before or at time $t$ (Group 1) and after time $t$ (Group 2). Let $n_{ik}(t)$ be the number of observations in group $i \ (i  = 1, 2)$ and category $k \ (k = 1,\ldotp\ldotp\ldotp,K)$ and $m_{k} \ (k = 1,\ldotp\ldotp\ldotp,K)$ be the number of observations in category $k$. Notice that $m_{k} = n_{1k}(t) + n_{2k}(t) \ (k = 1,\ldotp\ldotp\ldotp,K)$, $\sum_{k=1}^{K}m_{k} = n$, $\sum_{k=1}^{K}n_{1k}(t) = t$, and $\sum_{k=1}^{K}n_{2k}(t) = n-t$.
\begin{table*}[h]
	\centering
	\caption{Notations at time $t$}
	\label{t2}
	\begin{tabular}{cccccc}
		Index of distinct values & 1 & 2 & $\cdotp\cdotp\cdotp$ & $K$ & Total \\
		Group 1 & $n_{11}(t)$ & $n_{12}(t)$ & $\cdotp\cdotp\cdotp$ & $n_{1K}(t)$ & $t$ \\
		Group 2 & $n_{21}(t)$ & $n_{22}(t)$ & $\cdotp\cdotp\cdotp$ & $n_{2K}(t)$ & $n-t$  \\
		Total & $m_{1}$ & $m_{2}$ & $\cdotp\cdotp\cdotp$ & $m_{K}$ & $n$  \\
	\end{tabular}
\end{table*}

In \citet{chen2013graph} and \citet{zhang2017graph}, the authors studied ways to extend the underlying graph-based two-sample tests to accommodate data with repeated observations under the two-sample testing framework. When data has repeated observations, the similarity graph is not uniquely defined based on common optimization criteria, such as the MST, leading to multiple optimal graphs. The authors considered two ways to incorporate information from these graphs: averaging and union. To be more specific, they first construct the similarity graph on the distinct values, denoted by $C_{0}$. Here, $C_{0}$ could be the MST on all distinct values, the nearest neighbor link, the union of all possible MSTs on the distinct values, when the MST on the distinct values is not unique, or some other meaningful graphs. Then, the optimal graph initiated from $C_{0}$ is defined in the following way: For each pair of edges $(k_{1},k_{2}) \in C_{0}$, randomly choose an observation with the value indexed by $k_{1}$ and an observation with the value indexed by $k_{2}$, and connect these two observations; then, for each $k_{i} \ (i=1,2)$, if there are more than one observation (repeated observations) with the value indexed by $k_{i}$, connect them by a spanning tree (any edges in this spanning tree has distance 0). More detail explanations for $C_{0}$ are provided in Supplement B. Based on these optimal graphs, averaging statistic is defined by averaging the test statistic over all optimal graphs and union statistic is defined by calculating the test statistic on the union of all optimal graphs.


\section{Proposed tests} \label{sec:new}


\subsection{Extended test statistics for data with repeated observations} \label{sec:extended}

Here, we focus on extending the weighted, generalized, and max-type test statistics for repeated observations, which will turn out to be the asymptotic distribution-free tests. Details for extending the original edge-count test is in Supplement E. Based on the two-sample test statistics in \citet{chen2013graph} and \citet{zhang2017graph}, we could define the extended basic quantities at time $t$ under the averaging approach as follows:
\begin{align} 
R_{1,(a)}(t) &= \sum_{k=1}^{K}\frac{n_{1k}(t)\left(n_{1k}(t)-1\right)}{m_{k}} + \sum_{(u,v)\in C_{0}}\frac{n_{1u}(t)n_{1v}(t)}{m_{u}m_{v}}, \\
R_{2,(a)}(t) &= \sum_{k=1}^{K}\frac{n_{2k}(t)\left(n_{2k}(t)-1\right)}{m_{k}} + \sum_{(u,v)\in C_{0}}\frac{n_{2u}(t)n_{2v}(t)}{m_{u}m_{v}}, 
\end{align}
and under the union approach as follows:
\begin{align} 
R_{1,(u)}(t) &= \sum_{k=1}^{K}\frac{n_{1k}(t)\left(n_{1k}(t)-1\right)}{2}+ \sum_{(u,v)\in C_{0}}n_{1u}(t)n_{1v}(t), \\
R_{2,(u)}(t) &= \sum_{k=1}^{K}\frac{n_{2k}(t)\left(n_{2k}(t)-1\right)}{2}+ \sum_{(u,v)\in C_{0}}n_{2u}(t)n_{2v}(t).
\end{align}
These are discrete data version of $R_{1,G}(t)$ and $R_{2,G}(t)$ to address infeasibility of computing test statistics in data with repeated observations. Hence, relatively large value of the sum of $R_{1,(a)}(t)$ and $R_{2,(a)}(t)$ or $R_{1,(u)}(t)$ and $R_{2,(u)}(t)$ could be the evidence against the null hypothesis.

Under the null hypothesis $H_{0}$ (\ref{eq1.1}), the null distribution is defined to be the permutation distribution, which places $1/n!$ probabilities on each of the $n!$ permutations of $\{\by_{i}\}_{i=1,\ldotp\ldotp\ldotp,n}$. With no further specification, $\pr$, $\ep$, $\var$, and $\cov$ denote the probability, the expectation, the variance, and the covaraince, respectively, under the permutation null distribution. Then, analytic formulas for the expectation and the variance of extended basic quantities can be calculated through combinatorial analysis and detailed expressions and proof can be found in Supplement C.

For any candidate value $t$ of $\tau$, the extended test statistics can be defined as
\begin{alignat*}{2}
&\ \ Z_{w,(a)}(t) = \frac{R_{w,(a)}(t)-\ep\left(R_{w,(a)}(t)\right)}{\surd{\var\left(R_{w,(a)}(t)\right)}}, \quad &&Z_{w,(u)}(t) = \frac{R_{w,(u)}(t)-\ep\left(R_{w,(u)}(t)\right)}{\surd{\var\left(R_{w,(u)}(t)\right)}}, \\
& \ \ S_{(a)}(t) = Z_{w,(a)}^2(t) + Z_{d,(a)}^2(t), \quad &&S_{(u)}(t) = Z_{w,(u)}^2(t) + Z_{d,(u)}^2(t),\\
& \ \ M_{(a)}(t) = \max\left(Z_{w,(a)}(t),|Z_{d,(a)}(t)|\right), \quad
&&M_{(u)}(t) = \max\left(Z_{w,(u)}(t),|Z_{d,(u)}(t)|\right),
\end{alignat*}
where
\begin{align*}
R_{w,(a)}(t) = (1-w(t))R_{1,(a)}(t) + w(t)R_{2,(a)}(t),& \ \ R_{w,(u)}(t) = (1-w(t))R_{1,(u)}(t) + w(t)R_{2,(u)}(t), \\
Z_{d,(a)}(t) = \frac{R_{d,(a)}(t)-\ep\left(R_{d,(a)}(t)\right)}{\surd{\var\left(R_{d,(a)}(t)\right)}},& \ \ R_{d,(a)}(t) = R_{1,(a)}(t) - R_{2,(a)}(t), \\
Z_{d,(u)}(t) = \frac{R_{d,(u)}(t)-\ep\left(R_{d,(u)}(t)\right)}{\surd{\var\left(R_{d,(u)}(t)\right)}},&  \ \ R_{d,(u)}(t) = R_{1,(u)}(t) - R_{2,(u)}(t),
\end{align*}
with $w(t) = (t-1)/(n-2)$. Relatively large values of test statistics are the evidence against the null.

\subsection{New scan statistics}

Based on the extended statistics, we could define the scan statistics for the single change-point alternative to handle data with repeated observations as follows:
\begin{enumerate}
	\item Extended weighted edge-count scan statistic: $\max_{n_{0}\le t \le n_{1}}Z_{w,(a)}(t)$ \ \& \ $\max_{n_{0}\le t \le n_{1}}Z_{w,(u)}(t)$,
	\item Extended generalized  edge-count scan statistic: $\max_{n_{0}\le t \le n_{1}}S_{(a)}(t)$ \ \& \ $\max_{n_{0}\le t \le n_{1}}S_{(u)}(t)$,
	\item Extended max-type edge-count scan statistic: $\max_{n_{0}\le t \le n_{1}}M_{(a)}(t)$ \ \& \ $\max_{n_{0}\le t \le n_{1}}M_{(u)}(t)$,
\end{enumerate}
where $n_0$ and $n_1$ are set to be pre-specified values. For example, we can set $n_{0} = \lceil0.05n\rceil$ and $n_{1} = n-n_{0}$ in order to contain enough observations to represent the distribution.

Each scan statistic has its own characteristics aimed for different situations (see Section \ref{sec:simu} for a comparison of them). For example, the extended weighted edge-count test is useful when a change-point occurs away from the middle of the sequence. The extended generalized edge-count test is effective under both location and scale alternatives. The extended max-type edge-count test is similar but gives more accurate $p$-value approximation. The null hypothesis is rejected if the maxima is greater than a certain threshold. How to choose the threshold to control the type I error rate is described in Section \ref{sec:asym}.

\begin{figure}[h]
	\centering
	\includegraphics[width=4.4in]{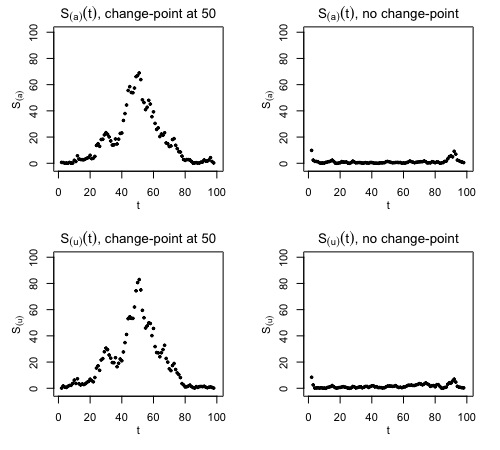}
	\caption{Plots of $S_{(a)}(t)$ and $S_{(u)}(t)$ against $t$ from a typical observation from Multinomial$(10, prob_{1})$ and the second 50 observations from Multinomial$(10, prob_{2})$ where $prob_{1} = (0.2,0.3,0.3,0.2)^{T}$ and $prob_{2}= (0.4,0.3,0.2,0.1)^{T}$ (left panel), and all 100 observations from Multinomial$(10, prob_{1})$ (right panel). Here, $C_{0}$ is the nearest neighbor link on Euclidean distance.} 
	\label{f2}
\end{figure}
For illustration, Figure \ref{f2} plots the processes of $S_{(a)}(t)$ and $S_{(u)}(t)$ for the first 50 observation generated from Multinomial$\left(10, (0.2,0.3,0.3,0.2)^{T}\right)$ and the second 50 observations generated from Multinomial$\left(10, (0.4,0.3,0.2,0.1)^{T}\right)$ with $C_0$ the nearest neighbor link constructed on the Euclidean distance. We see that both $S_{(a)}(t)$ and $S_{(u)}(t)$ peak at the true change-point $\tau = 50$ (left panel). On the other hand, when there is no change-point, $S_{(a)}(t)$ and $S_{(u)}(t)$ have random fluctuations with smaller maximum values (right panel). Illustrations of other test statistics are provided in Supplement D.

For testing the null $H_{0}$ (\ref{eq1.1}) against the changed-interval alternative $H_{2}$ (\ref{eq1.3}), test statistics can be derived in a similar way to the single change-point case. For example, the extended generalized edge-count scan statistics are
\begin{equation*}
	\max_{\substack{1<t_{1}<t_{2}\le n \\ n_{0}\le t_{2}-t_{1} \le n_{1}}}S_{(a)}(t_{1},t_{2}) \ \ \  \text{ and } \max_{\substack{1<t_{1}<t_{2}\le n \\ n_{0}\le t_{2}-t_{1} \le n_{1}}}S_{(u)}(t_{1},t_{2})
\end{equation*}
for the averaging and union approaches, respectively, where $S_{(a)}(t_{1},t_{2})$ and $S_{(u)}(t_{1},t_{2})$ are the extended generalized edge-count statistics on the two samples: observations within $[t_{1},t_{2})$ and observations outside $[t_{1},t_{2})$. The details of all statistics for the changed-interval alternative are in Supplement F.


\section{Analytical $p$-value approximation} \label{sec:asym}


\subsection{Asymptotic distributions of the stochastic processes} \label{s4.1}

{We first consider the 	averaging approach.}  We are concerned with the tail distribution of the scan statistics under $H_{0}$. Take the extended generalized edge-count scan statistic as an example, we want to compute
\begin{equation} \label{e4.1}
\pr\left(\max_{n_{0}\le t \le n_{1}}S_{(a)}(t)\right) , \ \ \pr\left(\max_{n_{0}\le t \le n_{1}}S_{(u)}(t)\right)
\end{equation}
for the single change-point alternative, and
\begin{equation} \label{e4.2}
\pr\left(\max_{\substack{1<t_{1}<t_{2}\le n \\ n_{0}\le t_{2}-t_{1} \le n_{1}}}S_{(a)}(t_{1},t_{2})\right) , \ \ \pr\left(\max_{\substack{1<t_{1}<t_{2}\le n \\ n_{0}\le t_{2}-t_{1} \le n_{1}}}S_{(u)}(t_{1},t_{2})\right)
\end{equation}
for the changed-interval alternative. 

Under the null hypothesis, the scan statistics are defined as the permutation distribution. For small sample size $n$, we can directly sample from the permutation distribution to compute the permutation $p$-value. However, when $n$ is large, one needs to draw a large number of random permutations to get a good estimate of the $p$-value, which is very time consuming. Hence, we seek to derive analytic approximations to these tail probabilities. 

By the definition of $Z_{w,(a)}(t)$, $S_{(a)}(t)$, and $M_{(a)}(t)$, stochastic processes $\{Z_{w,(a)}(t)\}$, $\{S_{(a)}(t)\}$, and $\{M_{(a)}(t)\}$ boil down to two pairs of basic processes: $\{Z_{w,(a)}(t)\}$ and $\{Z_{d,(a)}(t)\}$ for the single change-point case and $\{Z_{w,(a)}(t_{1},t_{2})\}$ and $\{Z_{d,(a)}(t_{1},t_{2})\}$ for the changed-interval case in the similar way. Therefore, we first study the properties of these basic stochastic processes. Let $\mathcal{E}_{u}^{C_{0}}$ be the subgraph of $C_{0}$ containing all edges that connect to node $u$, $\mathcal{E}_{u,2}$ be the set of edges in $C_{0}$ that contains at least one node in $\mathcal{E}_{u}^{C_{0}}$, and $|\mathcal{E}_{u}^{C_{0}}|$ and $|\mathcal{E}_{u,2}^{C_{0}}|$ be the number of edges in the sets. To derive the asymptotic behavior of the stochastic processes in averaging approach, we work under the following conditions:
\begin{condition} \label{c4.1}
	$|C_{0}|$, \ $\sum_{(u,v)\in C_{0}}(m_{u}m_{v})^{-1} = O(n)$.
\end{condition}
\begin{condition} \label{c4.2}
	$\sum_{u=1}^{K}m_{u}\left(m_{u}+|\mathcal{E}_{u}^{C_{0}}|\right)\left(\sum_{v\in\mathcal{V}_{u}^{C_{0}}}m_{v}+|\mathcal{E}_{u,2}^{C_{0}}|\right) = o(n^{3/2})$.
\end{condition}
\begin{condition} \label{c4.3}
	$\sum_{(u,v)\in C_{0}}\left(m_{u}+m_{v}+|\mathcal{E}_{u}^{C_{0}}|+|\mathcal{E}_{v}^{C_{0}}|\right)\big(\sum_{w\in\mathcal{V}_{u}^{C_{0}}\cup\mathcal{V}_{v}^{C_{0}}}m_{w}+|\mathcal{E}_{u,2}^{C_{0}}| + |\mathcal{E}_{v,2}^{C_{0}}|\big)$ \\   \indent\indent\indent\indent\indent\indent\indent\indent\indent $= o(n^{3/2})$.
\end{condition}
\begin{condition} \label{c4.4}
	$\sum_{u=1}^{K}(|\mathcal{E}_{u}^{C_{0}}|-2)^2/(4m_{u})-(|C_{0}|-K)^2/n = O(n)$.
\end{condition}

Let $[x]$ denotes the largest integer that is no larger than $x$.
\begin{theorem} \label{thm:averaging}
	Under Conditions \ref{c4.1}--\ref{c4.4}, as $n\rightarrow\infty$,
	\begin{enumerate}
		\item $\{Z_{w,(a)}([nw]) : 0 < w < 1\}$ and $\{Z_{d,(a)}([nw]) : 0 < w < 1\}$ converge to independent Gaussian processes in finite dimensional distributions, which we denote as $\{Z_{w,(a)}^{*}(w) : 0 < w < 1\}$ and $\{Z_{d,(a)}^{*}(w) : 0 < w < 1\}$, respectively.
		\item $\{Z_{w,(a)}([nw_{1}],[nw_{2}]) : 0 < w_{1} < w_{2} < 1\}$ and $\{Z_{d,(a)}([nw_{1}],[nw_{2}]) : 0 < w_{1} < w_{2} < 1\}$ converge to independent Gaussian random fields in finite dimensional distributions, which we denote as $\{Z_{w,(a)}^{*}(w_{1},w_{2}) : 0 < w_{1} <w_{2} < 1\}$ and $\{Z_{d,(a)}^{*}(w_{1},w_{2}) : 0 < w_{1} < w_{2} < 1\}$, respectively.
	\end{enumerate}
\end{theorem}

The proof for this theorem uses the technique developed in \cite{chen2015graph} that utilizes the Stein’s method \citep{ChenShao2005}. The details of the proof are in Supplement G.
%

Let $\rho_{w,(a)}^{*}(u,v) = \cov(Z_{w,(a)}^{*}(u),Z_{w,(a)}^{*}(v))$ and $\rho_{d,(a)}^{*}(u,v) = \cov(Z_{d,(a)}^{*}(u), Z_{d,(a)}^{*}(v))$. The next theorem states explicit expressions for the covariance functions of the limiting Gaussian process, $\{Z_{w,(a)}^{*}(w), \ 0<w<1\}$ and $\{Z_{d,(a)}^{*}(w), \ 0<w<1\}$.  It is proved through combinatorial analysis and details are given in Supplement H.

\begin{theorem} \label{t4.7}
	The covariance functions of the Gaussian processes $Z_{w,(a)}^{*}(w)$, and $Z_{d,(a)}^{*}(w)$ have the following expressions:
		\begin{align*}
		\rho_{w,(a)}^{*}(u,v) = \frac{(u\wedge v)\left\{1-(u\vee v)\right\} }{(u\vee v)\left\{ 1-(u\wedge v)\right\} }, \ \ 
		\rho_{d,(a)}^{*}(u,v) = \left[\frac{(u\wedge v)\left\{1-(u\vee v)\right\} }{(u\vee v)\left\{1-(u\wedge v)\right\} }\right]^{1/2},
		\end{align*}
	where $u\wedge v = \min(u,v)$ and $u\vee v = \max(u,v)$.
\end{theorem}




{For the union approach}, let $\bar{G}$ be the set of graphs that the union of all possible optimal graphs between observations $\{y_{i}\}$, $\mathcal{E}_{i}^{\bar{G}}$ be the subgraph of $\bar{G}$ containing all edges that connect to node $y_{i}$, and $|\mathcal{E}_{i}^{\bar{G}}|$ be the number of edges in the set. We work under
\begin{condition} \label{c4.8}
	$|\bar{G}| = O(n)$.
\end{condition}
\begin{condition} \label{c4.9}
	$\sum_{u=1}^{K}m_{u}^3\sum_{v\in\mathcal{V}_{u}^{C_{0}}}m_{v} \sum_{v\in\mathcal{V}_{u}^{C_{0}}}m_{v}\left(m_{v}+\sum_{w\in\mathcal{V}_{v}^{C_{0}}\backslash\{v\}}m_{w}\right)  = o(n^{3/2})$.
\end{condition}
\begin{condition} \label{c4.10}
	$\sum_{(u,v)\in C_{0}}m_{u}m_{v}\left(m_{u}\sum_{w\in\mathcal{V}_{u}^{C_{0}}}m_{w}+m_{v}\sum_{w\in\mathcal{V}_{v}^{C_{0}}}m_{w}\right)$  \\
	\indent\indent\indent\indent \indent\indent\indent\indent \indent\indent\indent $ \times \left\{\sum_{w\in\mathcal{V}_{u}^{C_{0}}\cup\mathcal{V}_{v}^{C_{0}}, \ y \in \mathcal{V}_{w}^{C_{0}}\backslash\{w\}}m_{w}\left(m_{w}+m_{y}\right)\right\} = o(n^{3/2}). $
\end{condition}
\begin{condition} \label{c4.11}
	$\sum_{i=1}^{n}|\mathcal{E}_{i}^{\bar{G}}|^2-4|\bar{G}|^2/n = O(n)$.
\end{condition}

\begin{theorem} \label{thm:union}
	Under Conditions \ref{c4.8}--\ref{c4.11}, as $n\rightarrow\infty$,
	\begin{enumerate}
		\item $\{Z_{w,(u)}([nw]) : 0 < w < 1\}$ and $\{Z_{d,(u)}([nw]) : 0 < w < 1\}$ converge to independent Gaussian processes in finite dimensional distributions, which we denote as $\{Z_{w,(u)}^{*}(w) : 0 < w < 1\}$ and $\{Z_{d,(u)}^{*}(w) : 0 < w < 1\}$, respectively.
		\item $\{Z_{w,(u)}([nw_{1}],[nw_{2}]) : 0 < w_{1} < w_{2} < 1\}$ and $\{Z_{d,(u)}([nw_{1}],[nw_{2}]) : 0 < w_{1} < w_{2} < 1\}$ converge to independent Gaussian random fields in finite dimensional distributions, which we denote as $\{Z_{w,(u)}^{*}(w_{1},w_{2}) : 0 < w_{1} <w_{2} < 1\}$ and $\{Z_{d,(u)}^{*}(w_{1},w_{2}) : 0 < w_{1} < w_{2} < 1\}$, respectively.
	\end{enumerate}
\end{theorem}
The details of the proof are in Supplement I.

Let $\rho_{w,(u)}^{*}(u,v) = \cov(Z_{w,(u)}^{*}(u),Z_{w,(u)}^{*}(v))$ and $\rho_{d,(u)}^{*}(u,v) = \cov(Z_{d,(u)}^{*}(u), Z_{d,(u)}^{*}(v))$. The next theorem states explicit expressions for the covariance functions of the limiting Gaussian processes, $\{Z_{w,(u)}^{*}(w), \ 0<w<1\}$, and $\{Z_{d,(u)}^{*}(w), \ 0<w<1\}$.  Its proof is in Supplement J.

\begin{theorem}  \label{t4.14}
	The covariance functions of the Gaussian processes $Z_{w,(u)}^{*}(w)$, and $Z_{d,(u)}^{*}(w)$ have the follwoing expressions:
	 \begin{align*}
		\rho_{w,(u)}^{*}(u,v) = \frac{(u\wedge v)\left\{1-(u\vee v)\right\} }{(u\vee v)\left\{ 1-(u\wedge v)\right\} }, \ \
		\rho_{d,(u)}^{*}(u,v) = \left[\frac{(u\wedge v)\left\{1-(u\vee v)\right\} }{(u\vee v)\left\{1-(u\wedge v)\right\} }\right]^{1/2}.
		\end{align*}
\end{theorem}

\begin{remark}
	By Theorems \ref{t4.7}, \ref{t4.14}, we see that the limiting distributions for $\{Z_{w,(a)}([nw]) : 0<w<1\}$ and $\{Z_{w,(u)}([nw]) : 0<w<1\}$ are the same and do not depend on the graph at all.  The same story for $Z_{d}$'s. In addition, their covariance functions in Theorem \ref{t4.7} and \ref{t4.14} are the same as Theorem 4.3 in \cite{chu2019asymptotic}. Hence, limiting distributions of the extended graph-based tests based on $Z_{w}$, $S$, $M$ are exactly the same as their corresponding versions for continuous data. On the other hand, the limiting distributions of the extended original edge-count scan statistics differ from their corresponding versions under the continuous setting (see details in Supplement E).
\end{remark}

\begin{remark} \label{reamark:averaging}
	Conditions \ref{c4.1}--\ref{c4.11} all constrain the number of repeated observations.  Conditions \ref{c4.1} and \ref{c4.8} can usually be satisfied with an appropriate choice of $C_{0}$. Conditions \ref{c4.2}, \ref{c4.3}, \ref{c4.9} and \ref{c4.10} constrain the degrees of nodes in the graph $C_{0}$ such that they cannot be too large. Condition \ref{c4.4} ensures that $(R_{1,(a)}(t),R_{2,(a)}(t))^{T}$ does not degenerate asymptotically so that $S_{(a)}(t)$ is well defined.  Similar for Condition \ref{c4.11}.
	
	{We check these conditions through simulation studies (Supplement L) and see that some of them could be violated even when the $p$-value approximation still works well.  
 \cite{zhu2021limiting} recently studied graph-based two-sample tests for continuous data and they proposed much more relaxed conditions than those in \cite{chu2019asymptotic}. They checked their conditions under both sparse and dense graphs and the conditions hold well.  We believe that conditions for data with repeated observations under the change-point setting can also be relaxed. This requires substantial work and we leave this to our future research.}

\end{remark}


\subsection{Asymptotic $p$-value approximation} \label{s4.3}
We now examine the asymptotic behavior of tail probabilities. Following similar arguments in the proof for Proposition 3.4 in \citet{chen2015graph}, we can obtain the foundation for analytical approximations to the probabilities. Assume that $n_{0}, n_{1}, n, b \rightarrow \infty$ in a way such that for some $0 < x_{0} < x_{1} < 1$ and $b_{1} > 0$, $n_{0}/n \rightarrow x_{0}, \ n_{1}/n \rightarrow x_{1}, \ b/\surd{n} \rightarrow b_{1}$.

Based on Theorem \ref{thm:averaging} and \ref{thm:union}, as $n\rightarrow\infty$, for both averaging and union approaches, we have
\begin{align*}
	\pr\left(\max_{n_{0}\leq t \leq n_{1}}Z_{w}^{*}(t/n) > b\right)  &\sim b\phi(b)\int_{x_{0}}^{x_{1}}h_{w}^{*}(x)\nu\left[b_{1}\left\{2h_{w}^{*}(x)\right\}^{1/2}\right]dx, \\
	\pr\left(\max_{n_{0}\leq t \leq n_{1}}|Z_{d}^{*}(t/n)| > b\right)  &\sim 2b\phi(b)\int_{x_{0}}^{x_{1}}h_{d}^{*}(x)\nu\left[b_{1}\left\{2h_{d}^{*}(x)\right\}^{1/2}\right]dx, 
\end{align*}
where $\nu(s) \approx (2/s)\left\{\Phi(s/2)-0.5\right\}/\left\{(s/2)\Phi(s/2)+\phi(s/2)\right\}$ \citep{siegmund2007statistics} with $\Phi(\cdot)$ and $\phi(\cdot)$ being the standard normal cumulative density function and probability density function, respectively, and
	\begin{align*}
	h_{w}^{*}(x) = \lim_{s\nearrow x}\frac{\partial\rho_{w,(a)}^{*}(s,x)}{\partial s} & = -\lim_{s\searrow x}\frac{\partial\rho_{w,(a)}^{*}(s,x)}{\partial s} =  \lim_{s\nearrow x}\frac{\partial\rho_{w,(u)}^{*}(s,x)}{\partial s}  = -\lim_{s\searrow x}\frac{\partial\rho_{w,(u)}^{*}(s,x)}{\partial s}, \\
	h_{d}^{*}(x) = \lim_{s\nearrow x}\frac{\partial\rho_{d,(a)}^{*}(s,x)}{\partial s} & = -\lim_{s\searrow x}\frac{\partial\rho_{d,(a)}^{*}(s,x)}{\partial s}=\lim_{s\nearrow x}\frac{\partial\rho_{d,(u)}^{*}(s,x)}{\partial s} = -\lim_{s\searrow x}\frac{\partial\rho_{d,(u)}^{*}(s,x)}{\partial s}.
	\end{align*}
It can be shown that $h_{w}^{*}(x)  = \left\{x(1-x)\right\}^{-1}$ and $h_{d}^{*}(x)  = \left\{2x(1-x)\right\}^{-1}$.

Since $Z_{w,(a)}^{*}(t)$ and $Z_{d,(a)}^{*}(t)$ are independent and $Z_{w,(u)}^{*}(t)$ and $Z_{d,(u)}^{*}(t)$ are independent, for both averaging and union approaches, we have
\begin{align*}
	\pr\left(\max_{n_{0}\leq t \leq n_{1}}M^{*}(t/n) > b\right) = 1-\pr\left(\max_{n_{0}\leq  t \leq n_{1}}|Z_{d}^{*}(t/n)| < b\right)\pr\left(\max_{n_{0}\leq t \leq n_{1}}Z_{w}^{*}(t/n) < b\right).
\end{align*}

In addition, following similar arguments in the proof for Proposition 4.4 in \citet{chu2019asymptotic}, we obtain the analytical $p$-value approximations for the extended generalized edge-count test. Assume that $n_{0}, n_{1}, n, b_{S} \rightarrow \infty$ in a way such that for some $0 < x_{0} < x_{1} < 1$ and $b_{2} > 0$, $n_{0}/n \rightarrow x_{0}, \ n_{1}/n \rightarrow x_{1}, \ b_{S}/n \rightarrow b_{2}$.
Then, as $n\rightarrow\infty$, for both averaging and union approaches, we have
	\begin{align*}
	&\pr\left(\max_{n_{0}\leq t \leq n_{1}}S^{*}(t/n) > b_{S}\right) \sim \frac{b_{S}e^{-b_{S}/2}}{2\pi}\int_{0}^{2\pi}\int_{x_{0}}^{x_{1}}h_{s}^{*}(x,w)\nu\left[\left\{2b_{2}h_{s}^{*}(x,w)\right\}^{1/2}\right]dxdw, 
	\end{align*}
where $h_{s}^{*}(x,w) = h_{d}^{*}(x)\cos^{2}(w) + h_{w}^{*}(x)\sin^{2}(w)$. The analytical $p$-value approximations for the changed-interval are in Supplement M.

\begin{remark}
	In practice, we use $h_{w}(n,x)$ in place of $h_{w}^{*}(x)$, where $h_{w}(n,x)$ is the finite-sample equivalent of $h_{w}^{*}(x)$. That is,
	\begin{equation*}
	h_{w}(n,x) = n\lim_{s\nearrow nx}\frac{\partial \rho_{w}(s,nx)}{\partial s},
	\end{equation*}
	with $\rho_{w}(s,t) = \cov\left(Z_{w}(s), Z_{w}(t)\right)$. The explicit expression for $h_{w}(n,x)$ is
	\begin{equation*}
	h_{w}(n,x) = \frac{(n-1)(2nx^2-2nx+1)}{2x(1-x)(n^2x^2-n^2x+n-1)}.
	\end{equation*}
	It is clear from above expression that $h_{w}(n,x)$ does not depend on the graph $C_{0}$ as well and it is easy to show that $\lim_{n\rightarrow\infty}h_{w}(n,x) = h_{w}^{*}(x)$. The finite-sample equivalent of $h_{d}^{*}(x)$ is exact the same as $h_{d}^{*}(x)$, that is,
	\begin{equation*}
	h_{d}(n,x) = n\lim_{s\nearrow nx}\frac{\partial \rho_{d}(s,nx)}{\partial s} = \frac{1}{2x(1-x)},
	\end{equation*}
	where $\rho_{d}(s,t) = \cov\left(Z_{d}(s), Z_{d}(t)\right)$. 
\end{remark}


\subsection{Skewness correction} \label{s4.4} 

The analytic $p$-value approximations based on asymptotic results give ballpark estimates of the $p$-values. However, they are in general not accurate enough if we set $n_{0}$ and $n_{1}$ close to the two ends and when the dimension is high (see the table in Section \ref{s4.5}). The inaccuracy is largely attributed to the fact that the convergence of $Z_{w,(a)}(t), Z_{w,(u)}(t), Z_{d,(a)}(t), Z_{d,(u)}(t)$ to the Gaussian distribution is slow when $t/n$ is close to 0 or 1. 

To improve the analytical $p$-value approximation, we add extra terms in the analytic formulas to correct for skewness. In our problem, the extent of the skewness depends on the value of $t$. Hence, we adopt a skewness correction approach discussed in \citet{chen2015graph} where different amount of the correction is done for different $t$:
In particular, this approach utilizes better approximation of the marginal probability by using the third moment, $\gamma$.

After skewness correction, the analytical $p$-value approximations for averaging approach are
\begin{align}
	\pr\left(\max_{n_{0}\leq t \leq n_{1}}Z_{w,(a)}(t) > b\right) \sim b\phi(b)\int_{n_{0}/n}^{n_{1}/n}H_{w,(a)}(nx)h_{w}(n,x)\nu\left[b\left\{2h_{w}(n,x)/n\right\}^{1/2}\right]dx, \label{e4.27} 
\end{align}
where 
\begin{equation*}
H_{w,(a)}(t) = \frac{\exp\big\{\frac{1}{2}(b-\hat{\theta}_{b,w,(a)}(t))^2+\frac{1}{6}\gamma_{w,(a)}(t)\hat{\theta}_{b,w,(a)}^{3}(t)\big\}}{\left\{1+\gamma_{w,(a)}(t)\hat{\theta}_{b,w,(a)}(t)\right\}^{1/2}}, \ \ \hat{\theta}_{b,w,(a)}(t) = \frac{\left\{1+2\gamma_{w,(a)}(t)b\right\}^{1/2}-1}{\gamma_{w,(a)}(t)},
\end{equation*}
\begin{align}
	&\pr\left(\max_{n_{0}\leq t \leq n_{1}}Z_{d,(a)}(t) > b\right)  \sim b\phi(b)\int_{n_{0}/n}^{n_{1}/n}H_{d,(a)}(nx)h_{d}(n,x)\nu\left[b\left\{2h_{d}(n,x)/n\right\}^{1/2}\right]dx, \label{e4.29} 
\end{align}
where 
\begin{equation*}
H_{d,(a)}(t) = \frac{\exp\big\{\frac{1}{2}(b-\hat{\theta}_{b,d,(a)}(t))^2+\frac{1}{6}\gamma_{d,(a)}(t)\hat{\theta}_{b,d,(a)}^{3}(t)\big\}}{\left\{1+\gamma_{d,(a)}(t)\hat{\theta}_{b,d,(a)}(t)\right\}^{1/2}}, \ \ \hat{\theta}_{b,d,(a)}(t) = \frac{\left\{1+2\gamma_{d,(a)}(t)b\right\}^{1/2}-1}{\gamma_{d,(a)}(t)},
\end{equation*}
and $\gamma_{w,(a)}(t) =\ep\left(Z^3_{w}(t)\right), \gamma_{d,(a)}(t)= \ep\left(Z^3_{d}(t)\right)$, whose analytic expressions are provided in Supplement K. The skewness corrected analytical $p$-value approximations for union approach and the changed-interval can be derived in the similar manner and details are provided in Supplement M.

\begin{remark}
	By jointly correcting for the marginal probabilities of $Z_{w}(t)$ and $Z_{d}(t)$, we can derive skewness corrected $p$-value approximations for $\max_{n_{0}\le t \le n_{1}}S(t) = \max_{0\le w \le 2\pi}\max_{n_{0}\le t \le n_{1}}\left\{Z_{w}(t)\sin(w)+Z_{d}(t)\cos(w)\right\}$  \citep{chu2019asymptotic}. However, the integrand could be easily non-finite, so the method heavily relies on extrapolation. We thus do not perform skewness correction on $S_{(a)}(t)$ and $S_{(u)}(t)$.
\end{remark}


\subsection{Checking analytical $p$-value approximations under finite $n$} \label{s4.5}

We check the performance of analytical $p$-value approximations obtained in Section \ref{s4.3} and \ref{s4.4}. In particular, we compare the critical values for 0.05 $p$-value threshold through analytical $p$-value approximations based on asymptotic results and skewness correction to those obtained from doing 10,000 permutations under various simulation settings to check how analytical approximation works well for finite samples. Here, we focus on the extended max-type scan statistic for the single change-point alternative.  The results for other scan statistics and the changed-interval alternative are provided in Supplement N. 

We consider three distributions with different dimensions (Multinomial $d=10$ with equal probabilities (C1), Gaussian with repeated observations $d = 100$ (C2), Multinomial $d=1000$ with equal probabilities (C3)) and let $C_{0}$ be the nearest neighbor link constructed on Euclidean distance. The analytic approximations depend on constraints, $n_{0}$ and $n_{1}$, on the region where the change-point is searched. To make things simple, we set $n_{1} = n - n_{0}$.

Since analytical $p$-value approximations without skewness correction do not depend on $C_{0}$ in the extended weighted, generalized, and max-type tests, the critical value is determined by $n$, $n_{0}$, and $n_{1}$ only. Notice that analytical $p$-value approximations without skewness correction provide the same result in both averaging and union approaches. On the other hand, the skewness corrected approximated $p$-values depend on certain characteristics of the graph structure. The structure of the nearest neighbor link depends on the underlying dataset, and thus the critical values vary by simulation runs.

Table \ref{t6} shows results of the extended max-type scan statistics. The first table labeled `A1' presents the analytical critical values without skewness correction. `A2 (a)' and `A2 (u)' represent skewness corrected analytical critical values in averaging and union approaches, respectively, and `Per $(a)$' and `Per $(u)$' represent permutation critical values in averaging and union cases, respectively. We also show results for 2 randomly simulated sequences in each setting. We see that the asymptotic $p$-value approximation is doing reasonably well. As window size decreases, the analytical critical values become less precise. However, skewness corrected approximation performs much better than the approximation without skewness correction. When the dimension is not too high, such as (C1), the skewness corrected analytical approximation is doing reasonably well for $n_{0}$ as low as 25. When the dimension is high, such as (C2) and (C3), the approximation performs well when $n_{0} \ge 50$.
\begin{table}[htp!]
	\centering 
	\caption{Critical values for the single change-point scan statistic $\max_{n_0 \le t \le n_1} M_{(a)}(t)$ and $\max_{n_0 \le t \le n_1} M_{(u)}(t)$ based on the nearest neighbor link at 0.05 significance level. $n = 1000$}
	\label{t6}
	\begin{tabular}{ccccc}
		& $n_0 = 100$ & $n_0 = 75$ & $n_0 = 50$ & $n_0 = 25$\\
		A1  & 3.24 & 3.28 & 3.32 & 3.38\\
	\end{tabular} 
	
	\vspace{3mm}
	
	\begin{tabular}{p{0.7cm}cccccccc} 
		& \multicolumn{8}{c}{Critical Values $(a)$}  \\
		& \multicolumn{2}{c}{$n_0 = 100$} &  \multicolumn{2}{c}{$n_0 = 75$} &  \multicolumn{2}{c}{$n_0 = 50$} &  \multicolumn{2}{c}{$n_0 = 25$} \\
		& A2 $(a)$ &  Per $(a)$ & A2 $(a)$ & Per $(a)$ &  A2 $(a)$  & Per $(a)$ & A2 $(a)$ & Per $(a)$ \\
		\multirow{2}{1.6cm}{(C1)}
		& 3.30 & 3.30 & 3.36 & 3.37 & 3.43 & 3.43 & 3.54 & 3.58\\
		& 3.30 & 3.30 & 3.35 & 3.36 &  3.43 & 3.46 & 3.55 & 3.62\\
		\multirow{2}{1.6cm}{(C2)}
		& 3.36 & 3.34 & 3.44 & 3.45 & 3.56 & 3.59 & 3.72 & 3.98\\
		& 3.34 & 3.36 & 3.42 & 3.47 & 3.53 & 3.64 & 3.76 & 4.03\\
		\multirow{2}{1.6cm}{(C3)}
		& 3.30 & 3.30 & 3.38 & 3.41 & 3.48 & 3.57 & 3.67 & 3.93\\
		& 3.30 & 3.28 & 3.38 & 3.39 & 3.48 & 3.56 & 3.67 & 3.87\\
	\end{tabular}    
	
	\vspace{3mm}
	
	\begin{tabular}{p{0.7cm}cccccccc} 
		& \multicolumn{8}{c}{Critical Values $(u)$}  \\
		& \multicolumn{2}{c}{$n_0 = 100$} &  \multicolumn{2}{c}{$n_0 = 75$} &  \multicolumn{2}{c}{$n_0 = 50$} &  \multicolumn{2}{c}{$n_0 = 25$} \\
		& A2 $(u)$ &  Per $(u)$ & A2 $(u)$ & Per $(u)$ &  A2 $(u)$  & Per $(u)$ & A2 $(u)$ & Per $(u)$ \\
		\multirow{2}{1.6cm}{(C1)}
		& 3.32 & 3.30 & 3.37 & 3.40 & 3.44 & 3.43 & 3.54 & 3.59\\
		& 3.31 & 3.32 & 3.36 & 3.35 & 3.43 &  3.46 & 3.55 & 3.63\\
		\multirow{2}{1.6cm}{(C2)}
		& 3.35 & 3.36 & 3.42 & 3.43 & 3.51 & 3.52 & 3.62 & 3.80\\
		& 3.34 & 3.39 & 3.40 & 3.46 & 3.48 & 3.55 & 3.67 & 3.84\\
		\multirow{2}{1.6cm}{(C3)}
		& 3.31 & 3.30 & 3.39 & 3.41 & 3.50 & 3.57 & 3.69 & 3.93\\
		& 3.31 & 3.28 & 3.39 & 3.39 & 3.50 & 3.56 & 3.69 & 3.87\\
	\end{tabular} 
\end{table} 

\vspace*{-0.4cm}


\section{Performance of the new tests} \label{sec:simu}

We study the performance of the new tests in two aspects: (1) whether the test can reject the null hypothesis of homogenity when there is a change, and (2) if the test can reject $H_{0}$, whether the test can estimate the location of the change-point accurately. We use the configuration model random graph $G(v,\overrightarrow{k})$ to generate networks. Here, $v$ is the number of vertices and $\overrightarrow{k} = (k_{1}, \ldotp\ldotp\ldotp,k_{v})$ is a degree sequence on $v$ vertices, with $k_{i}$ being the degree of vertex $i$. To generate configuration model random graphs, given a degree sequence, we choose a uniformly random matching on the degree stubs (half edges).

We generate a sequence of $n = 200$ networks from the following model:
\begin{equation*} 
y_{i} \sim \left\{
\begin{tabular}{c} 
$G(v,\overrightarrow{k}_{1})$,  \ $i = 1,\ldotp\ldotp\ldotp,\tau$; \ \ \ \ \ \ \ \ \ \ \ \\
$G(v,\overrightarrow{k}_{2})$, \ $i = \tau+1,\ldotp\ldotp\ldotp,200$. 
\end{tabular}
\right.
\end{equation*}
We explore two cases of the location of the change-point, in the middle $(\tau = 100)$ and close to one end $(\tau = 170)$ for $v=6$ vertices in this simulation. This dataset has repeated networks. Also, we consider two types of changes: 
\begin{enumerate}
	\item An equal degree changes in the network (all elements in $\overrightarrow{k}_{1}$ are 2 and 2 elements in $\overrightarrow{k}_{2}$ are 4 and the rest are 2), 
	\item A random degree changes in the network (all elements in $\overrightarrow{k}_{1}$ are 2 and 2 elements in $\overrightarrow{k}_{2}$ are randomly selected from 3 to 5 and the rest are 2).
\end{enumerate}
That is, we present the results for the four combinations: an equal degree change at $\tau=100$ (S1), a random degree change at $\tau=100$ (S2), an equal degree change at $\tau=170$ (S3), and a random degree change  at $\tau=170$ (S4).

For network at $t$, we use an adjacency matrix $M_{t}$ to represent the network, with 1 for element $(i,j)$ if vertex $i$ and $j$ are connected, and 0 otherwise. We consider the dissimilarity defined as the number of different entries normalized by the geometric mean of the total edges in each of two networks, $\|M_{i}-M_{j}\|_{F}/\left(\|M_{i}\|_{F}\cdot\|M_{j}\|_{F}\right)^{1/2}$, where $\|\cdot\|_{F}$ is the Frobenius norm of a matrix. We set $C_{0}$ to be the nearest neighbor link as a similarity graph for our new methods.

\begin{table*} [htp!]
	\caption{Estimated power of new tests}
	\centering
	\label{t7}
	\begin{tabular}{ccccccc}
		&$Z_{w,(a)}(t)$ & $Z_{w,(u)}(t)$ & $S_{(a)}(t)$ & $S_{(u)}(t)$ & $M_{(a)}(t)$ & $M_{(u)}(t)$ \\
		(S1) & 0.98 (0.96) & 0.96 (0.89) & 0.96 (0.94)& 0.95 (0.89) &0.96 (0.95) & 0.95 (0.88) \\
		(S2) & 0.88 (0.83) & 0.89 (0.85) & 0.90 (0.84)& 0.91 (0.85) &0.89 (0.83) & 0.90 (0.87) \\
		(S3) & 0.86 (0.83) & 0.65 (0.59) & 0.85 (0.83)& 0.85 (0.82) &0.81 (0.80) & 0.70 (0.64) \\
		(S4) & 0.81 (0.81) & 0.73 (0.70) & 0.86 (0.84)& 0.93 (0.91) &0.84 (0.81) & 0.86 (0.82) \\
	\end{tabular}
\end{table*}

Table \ref{t7} shows the number of null rejection, out of 100, at 0.05 significance level for each method. For the accuracy of estimating the location of change-point, the count where the estimated change-point is within 20 from the true change-point is provided in parentheses when the null hypothesis is rejected. We see that all tests work well in the balanced equal degree changes case, while the extended generalized edge-count test outperforms in random degree changes case. In this simulation, equal degree changes would be considered the mean change and random degree changes would be considered the change in both location and scale. Hence, the extended generalized edge-count test and max-type edge-count test perform well in this general scenario. When the change-point is not in the center of the sequence, the extended weighted edge-count test outperforms, which complies with what we would expect. We see that the extended generalized edge-count test and max-type edge-count test work well when the change is in both mean and variance in the unbalnced sample size case.


\section{Phone-call network data analysis} \label{sec:phone}

Here, we apply the new tests to the phone-call network dataset mentioned in Section \ref{sec:int} in details. The MIT Media Laboratory conducted a study following 87 subjects who used mobile phones with a pre-installed device that can record call logs. The study lasted for 330 days from July 2004 to June 2005 \citep{eagle2009inferring}. Given the richness of this dataset, one question of interest to answer is that whether there is any change in the phone-call pattern among subjects over time. This can be viewed as the change of friendship along time.

We bin the phone-calls by day and we construct $t=330$ of networks in total with 87 subjects as nodes. We encode each network by the adjacency matrix $B_{t}$ with value 1 for element $(i,j)$ if subject $i$ called $j$ on day $t$ and 0 otherwise. We define the distance measure as the number of different entries, i.e., $d(B_{i}, B_{j}) = \|B_{i}-B_{j}\|_{F}^2$, where$\|\cdot\|_{F}$ is the Frobenius norm of a matrix. Due to the repeated observations, many equal distances among distinct values exist. We set $C_{0}$ to be the nearest neighbor link in this example.

We apply the single change-point detection method using the extended generalized scan statistic to the phone-call network dataset recursively so as to detect all possible change-points. As this dataset has a lot of noise, we focus on estimated change-points with $p$-value less than 0.001. Figure 2 shows estimated change-points by averaging approach and union approach. {We see that the two approaches produce quite a number of similar change-points.  We define a change-point $\hat{\tau}$ to be detected by both approaches if they each finds a change-point within the set $[\hat{\tau}-2, \hat{\tau}+2]$.  We then deem the location of the shared change-point to be the floor of the average of the two change-points detected by the two approaches.} 

Since we do not know the underlying distribution of the dataset, we perform more sanity check with the distance matrix of the whole period (Figure \ref{heatmap}). It is evident that there are some signals in this dataset and they show comparably good match with our results from the new tests.

\begin{figure}[htp!]
	\centering
	\includegraphics[width=4.2in]{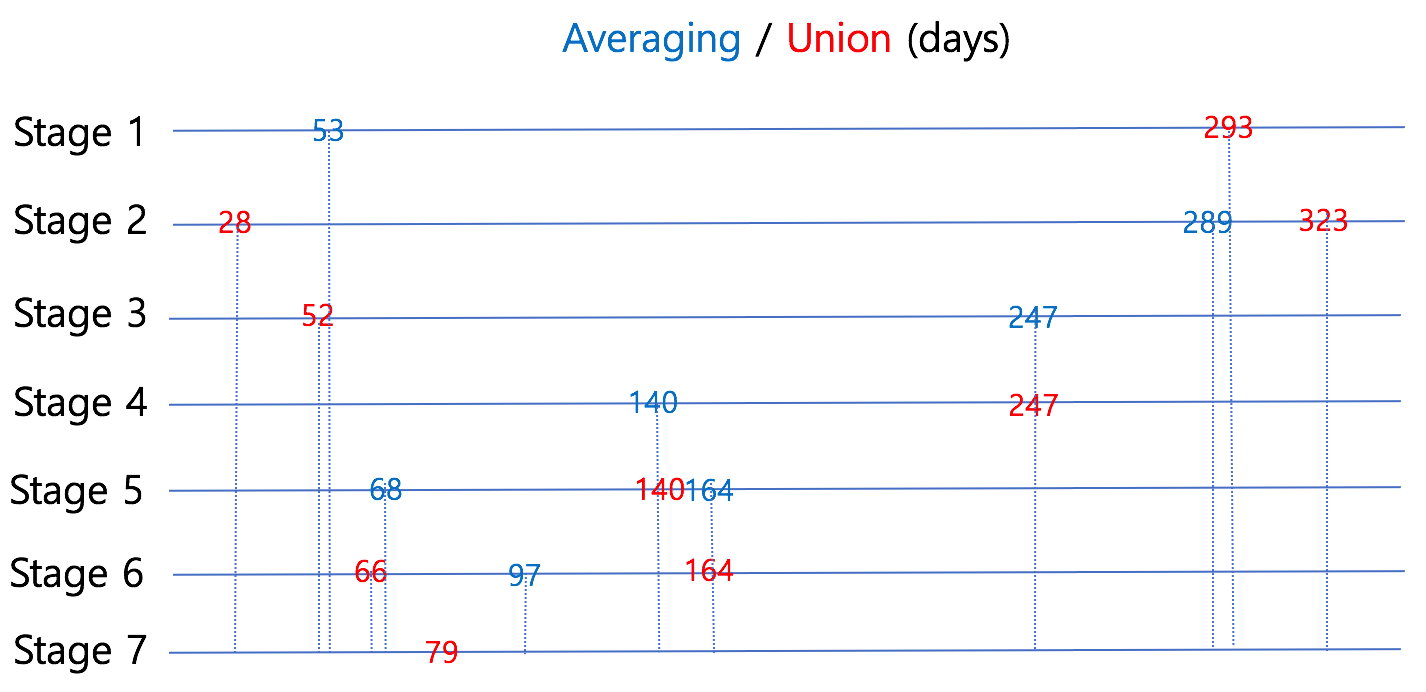} 
	\\ \ \\ 
	\begin{tabular}{cccccccccc}
		(a) &53 & 68 & 97 & 140 & 164 & 247 & 289 & & \\
		(u) & 28 & 52 & 66 & 79 &140 & 164 & 247 & 293 & 323 \\
		Shared & 52  & 67 & 140 & 164 & 247 & 291   & & & \\
	\end{tabular}
    \label{phone1}
	\caption{Estimated change-points and the order where change-points are detected for averaging and union approaches.} 
\end{figure}

\begin{figure}[htp!]
	\centering
	\includegraphics[width=2.9in]{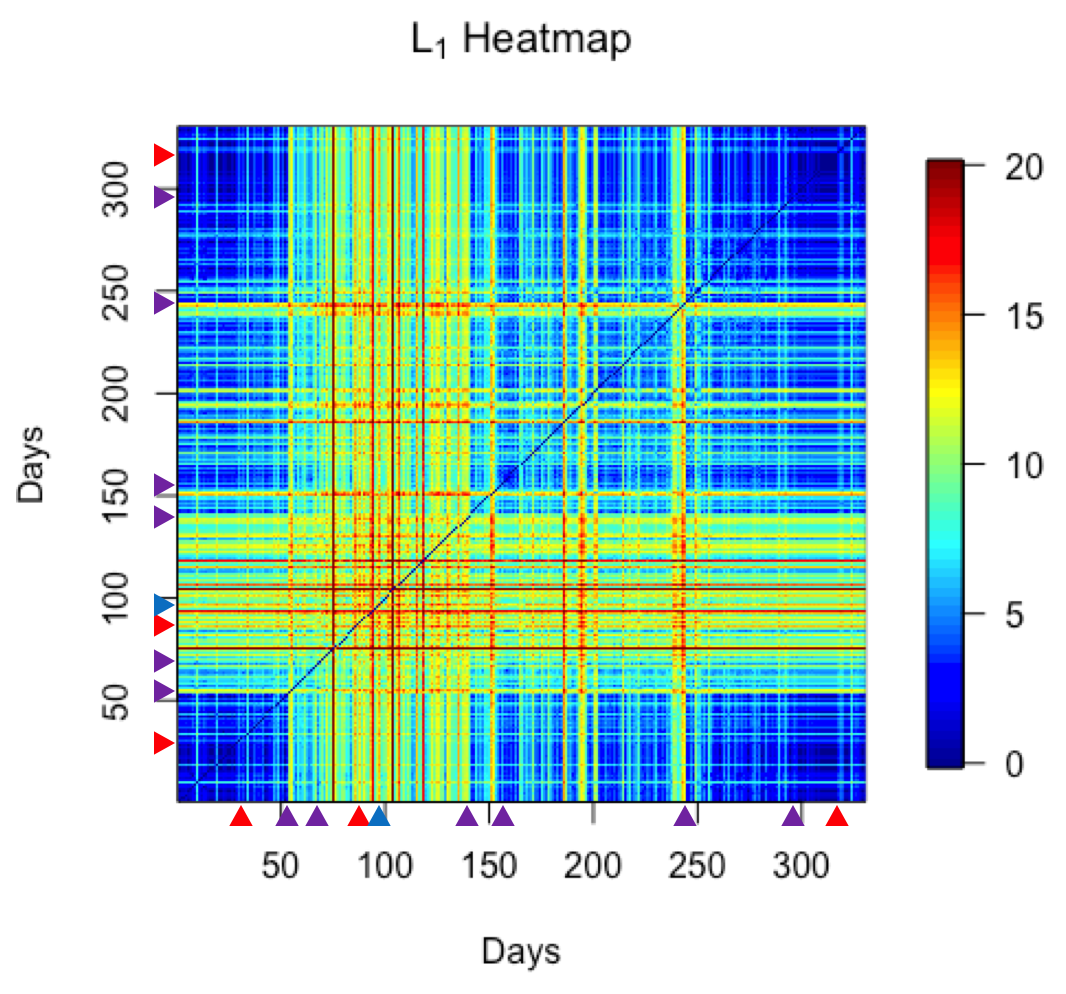}
	\caption{Heatmap of $L_{1}$ norm distance matrix corresponding to 330 networks. Red triangles, blue triangles, and purple triangles indicate estimated change-points by union approach, averaging approach, and their shared change-points, respectively.} 
	\label{heatmap}
\end{figure}


\vspace*{-0.7cm}

\section{Discussion}

{In general, the two approaches work similarly; while they inevitably produce different results sometimes.  A brief comparison of the two approaches is provided in Supplement O}. 
The proposed methods detect the most significant single change-point or the changed-interval in the sequence.   If the sequence has more than one change-point, the proposed methods can be applied recursively using techniques, such as binary segmentation, circular binary segmentation, or wild binary segmentation \citep{vostrikova1981detecting, olshen2004circular, fryzlewicz2014wild}.


\section*{Acknowledgement}
Hoseung Song and Hao Chen are supported in part by NSF awards DMS-1513653 and DMS-1848579.



%
%



\bibliographystyle{biometrika}
\bibliography{paper-ref}

\end{document}